\documentclass[11.5pt]{article}
\usepackage[top=1.5in, bottom=1.5in, left=1.2in, right=1.2in]{geometry}

\usepackage{latexopts}
\usepackage[sort]{natbib}
\usepackage{mathtools}\mathtoolsset{showonlyrefs,showmanualtags}
\usepackage{appendix}

\newcommand{\Hash}{\SC{Hash}\xspace}
\newcommand{\Reduce}{\SC{Reduce}\xspace}
\newcommand{\ReduceMSet}{\SC{ReduceMSet}\xspace}

\newcommand{\Rehash}{\SC{Rehash}\xspace}
\newcommand{\NullHash}{\ensuremath{\emptyset}\xspace}
\newcommand{\Union}{\SC{Union}\xspace}
\newcommand{\Intersection}{\SC{Intersection}\xspace}
\newcommand{\Difference}{\SC{Difference}\xspace}
\newcommand{\MarkerUnion}{\SC{MarkerUnion}\xspace}
\newcommand{\MarkerIntersection}{\SC{MarkerIntersection}\xspace}
\newcommand{\Snapshot}{\SC{Snapshot}\xspace}
\newcommand{\Summarize}{\SC{Summarize}\xspace}
\newcommand{\appsection}[1]{\let\oldthesection\thesection
  \renewcommand{\thesection}{Appendix \oldthesection}
  \section{#1}\let\thesection\oldthesection}

\usepackage{setspace}

\title{Efficient Identification of Equivalences in Dynamic Graphs and Pedigree Structures}

\author{
\mbox{
\begin{centering}
\begin{tabular}{c}
  Hoyt Koepke and Elizabeth Thompson\footnote{Hoyt Koepke (E-mail: \texttt{hoytak@stat.washington.edu}) is corresponding author and Graduate Student at University of Washington, Department of Statistics.  Elizabeth Thompson (E-mail: \texttt{eathomp@u.washington.edu}) is Professor at University of Washington, Department of Statistics, 
Seattle, WA 98195-4322.
 This work was supported in part by NIH grant GM-46255.} \\
\end{tabular}
\end{centering}
}}

\begin{document}

\maketitle

\begin{abstract}
  We propose a new framework for designing test and query functions for complex structures that vary across a given parameter such as genetic marker position.  The operations we are interested in include equality testing, set operations, isolating unique states, duplication counting, or finding equivalence classes under identifiability constraints.  A motivating application is locating equivalence classes in identity-by-descent (IBD) graphs, graph structures in pedigree analysis that change over genetic marker location.  The nodes of these graphs are unlabeled and identified only by their connecting edges, a constraint easily handled by our approach.  The general framework introduced is powerful enough to build a range of testing functions for IBD graphs, dynamic populations, and other structures using a minimal set of operations. The theoretical and algorithmic properties of our approach are analyzed and proved. Computational results on several simulations demonstrate the effectiveness of our approach.

  \vspace{.5 cm} \noindent {\bf Keywords: Hash Functions, Algorithms, Data Structures,
    Pedigree Analysis, Identity-by-Descent Graphs}
\end{abstract}

\newpage 

\section{Introduction}
\label{sec:intro}

In modern genetic analyses, we have genetic marker data $\mY_M$ on
individuals available at multiple genetic markers across the genome,
and the parameters of genetic marker models, $\mGamma_M$, are generally well
established. By contrast the models $\mGamma_T$ underlying trait data are less
clear. The goal of genetic linkage analyses is to locate DNA that
affects the trait relative to the known locations of the genetic
marker map. This requires computation of the conditional probability
$\Pof{\mY_T \C \mY_M \,;\, \mGamma}$, where $\mGamma$ is a joint model comprising $\mGamma_M$ and $\mGamma_T$ and a
specification $\gamma$ of the relative genome locations of DNA underlying $\mY_M$
and $\mY_T$. This probability will be required for multiple specifications
of the location of DNA affecting the trait, and may be required for
multiple values of trait parameter values $\mGamma_T$ and potentially for
multiple trait phenotypes on the same sets of related individuals.

This key probability is most easily considered as
\begin{equation}
  \Pof{\mY_T \C \mY_M \,;\, \mGamma} 
  = \sum_{\mZ} \Pof{\mY_T \C \mZ \,;\, \gamma, \mGamma} 
  \Pof{\mZ \C \mY_M \,;\, \mGamma_M}
  \label{eq:41AT}
\end{equation}
where $\mZ$ is an collection of latent variables insuring the
conditional independence of $\mY_M$ and $\mY_T$ given
$\mZ$. Classically, the chosen latent variables $\mZ$ were the
unobserved genotypes (types of the DNA) of the individuals of the
pedigree structure \citep{elston1971general, lathrop1984strategies}.
More recently a specification of the inheritance of the DNA at all
relevant locations has been the preferred choice of $\mZ$
\citep{lander1987construction,lange1991random,thompson1994monte-2}. On
large pedigrees, with data at multiple marker locations, the
probability in \eqref{eq:41AT} cannot be computed exactly, especially
if the data are sparse on the pedigree structures or if these
structures are complex. Instead, realizations of $\mZ$ from $\Pof{\mZ
  \C \mY_M \,;\, \mGamma_M}$ are obtained.  A Monte Carlo estimate of
$\Pof{\mY_T \C \mY_M \,;\, \mGamma}$ is the mean of the values of
$\Pof{\mY_T \C S^{(k)}\,;\, \gamma, \mGamma_T}$ over the $N$ realized
$\Set{\mZ^{(k)}}{k = 1,..., N}$.  Several effective MCMC methods have
been developed to obtain these realizations
\citep{sobel1996descent,thompson2000statistical,tong2008multilocus}.

In the context of modern informative marker data, an efficient choice
of $\mZ$ is the pattern of gene identity by descent (IBD), across the
chromosome, among individuals observed for the trait
\citep{thompson2011structure}. This defines a graph, the IBD graph,
which, at each locus, is analogous to the descent graph of
\citep{sobel1996descent}. The edges of this graph are the observed
individuals, and the nodes represent IBD sharing at this genome
location among the edges (individuals) connecting to that node.  The
IBD graph is a deterministic function of the inheritance
specification, and for marker genotypes $Y_M$ observed without error,
computation of the probability of these data for a given IBD graph is
easy \citep{sobel1996descent,kruglyak1996parametric}, Thus use of the
IBD graph led to greater efficiencies in obtaining MCMC realizations
from $\Pof{\mZ \C \mY_M \,;\, \mGamma_M}$.  However, it has been less
well appreciated that computation of $\Pof{\mY_T \C \mZ^{(k)}\,;\,
  \gamma, \mGamma_T}$ is also straightforward using the IBD-graph
representation \citep{thompson1999estimation,thompson2003}.

Use of the IBD-graph has other immediate advantages. They are %
generally slowly varying across the chromosome, relative to modern
marker densities, and may be output from the MCMC in compact format,
with only the change points and changes specified. Once the IBD graph
is realized, the pedigree structure is no longer required in
subsequent trait-data analyses, providing greater data
confidentiality, and the same set of realized IBD graphs may be used
for multiple values of $(\gamma, \mGamma_T)$ and even multiple
different traits observed on the same or different subsets of the
individuals \citep{thompson2011structure}. When reduced to the subset
of individuals observed for a trait, components of an IBD-graph $\mZ$ are
generally small, so that for single-locus models $\Pof{\mY_T \C
  \mZ\,;\, \gamma, \mGamma_T}$ is very easily computed, In fact,
computation on the joint graphs at several genome locations is also
feasible, leading to methods for genetic analysis under oligogenic
models \citep{su2012computationally}. Finally, the IBD framework has a
key advantage in that it is not dependent on the source of the
inferred IBD. 

Using population-based methods, IBD may be inferred between any two
individuals not known to be related
\citep{browning2010high,brown2012inferring}. If these individuals are
pedigree founders or members of different pedigrees, such
population-based IBD may be combined with pedigree-based inferences of
IBD to create merged IBD graphs, provided greater power and resolution
to trait analyses \citep{glazner2012}.

There is another huge computational advantage 
potentially available from the IBD-graph framework. In an IBD graph,
nodes have an identity only through the edges that connect them. Many
different inheritance patterns S give rise to the same node-unlabelled
IBD graph on the subset of trait-observed individuals. In an MCMC
analysis, many different realizations of $\mZ$ from $\Pof{\mZ \C \mY_M
  \,;\, \mGamma_M}$ may give the same IBD graph. Additionally, because
IBD-graphs are generally slowly varying, a given realized IBD-graph
may remain constant over several Mbp. Clearly, $\Pof{\mY_T \C \mZ\,;\,
  \gamma, \mGamma_T}$ should be computed once only for each distinct
$\mZ$. Recognition of when IBD-graphs are equal and of the marker ranges
over which they are equal, is crucial to efficient trait-data
analyses. The software developed in this paper performs this task
efficiently, and can decrease the burden of the trait-data probability
portion of the LOD score estimation procedure by up to two orders of
magnitude in real studies \citep{marchani2011estimation}. %


The key of our approach is to represent the object properties relevant to the testing by sets of representative hashes instead of the objects themselves.  Hashes permit much faster algorithms in many cases; for example, testing whether two graphs are equal can be done by checking whether two hashes are equal. These hashes are strong in the sense that intersections -- unequal objects or processes mapping to identical hashes -- are so unlikely as to never occur in practice.  Furthermore, we introduce several provably strong operations on such hashes that allow accurate reductions of collections while maintaining specified relationships between the hashes.  Thus in our approach, designing test functions is equivalent to designing composite hash functions accepting one or more input objects and returning a representative hash.  Testing equality over collections of input objects is then equivalent to testing equality of the output hashes; set operations over object collections is equivalent to set operations over the hashes, and so on.

We allow the objects in our framework to change over an indexing parameter.  We refer to this index as a {\it marker}, as it refers to genetic marker position in our target application, but it could just as easily refer to time or any other indexing parameter.  The power of this framework is that the building block operations process along all the possible marker values; the difficulties introduced by dynamic data is abstracted away.

\begin{figure}[tb]
  \begin{centering}
    \centerline{%
      \hss
      \subfloat[\label{fig:example-graph:a} $G(m_1)$] 
      {\includegraphics[width=0.3\linewidth]{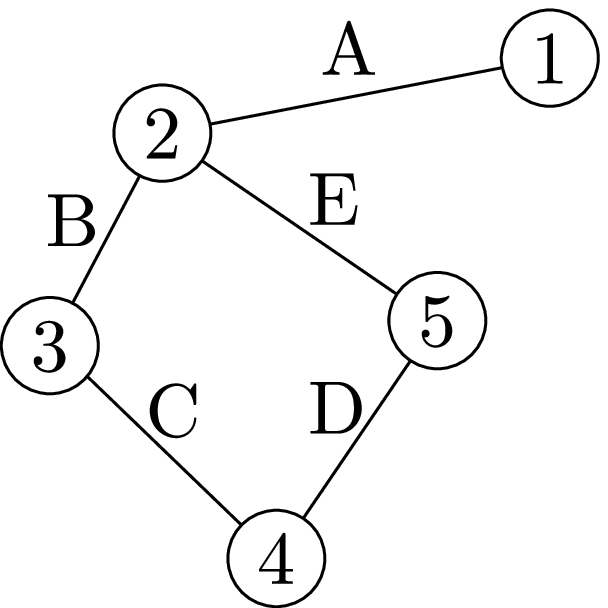}}
      \hss  \subfloat[\label{fig:example-graph:b} $G(m_2)$] 
      {\includegraphics[width=0.3\linewidth]{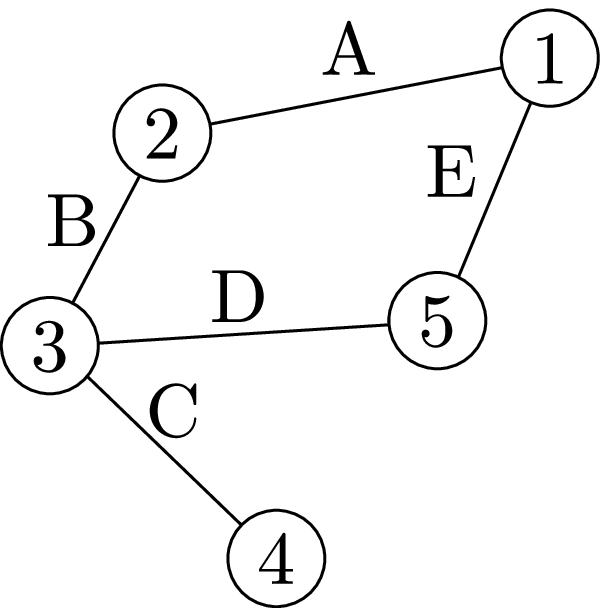}}
      \hss  \subfloat[\label{fig:example-graph:c} $G(m_3)$] 
      {\includegraphics[width=0.3\linewidth]{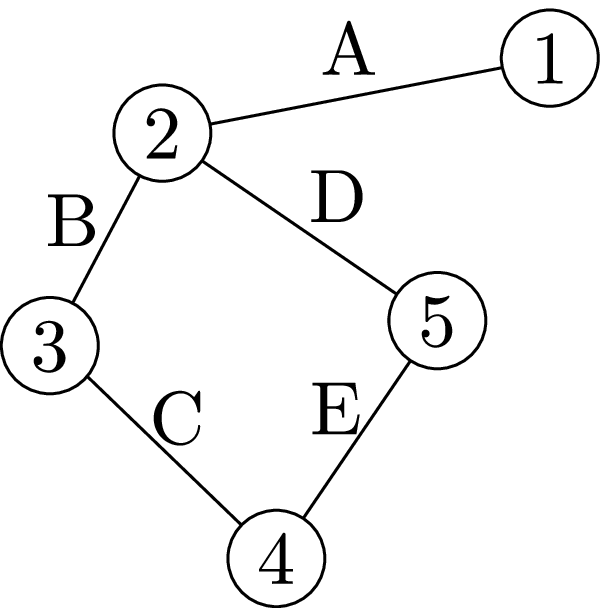}}
      \hss
    }
  \end{centering}
  \caption{\small{An example IBD graph $G[m]$ with labels on both
      edges and nodes.  The nodes (numbered) represent genetic
      sequences, while the edges (lettered) represent individuals.
      The graph changes slightly by marker value and is shown at
      marker values $m_1$, $m_2$, and $m_3$. Note that under the
      identifiability constraints for IBD graphs, (a) and (c) are
      distinct despite having the same skeletal structure.}}
  \label{fig:example-graph}
\end{figure}

The running example we use to illustrate this framework are
identity-by-descent graphs, or {\it IBD graphs}
\citep{sobel1996descent,ThompsonHeath99}. As an abstracted structure,
these graphs have two interesting and distinctive properties.  First,
only the links are identifiable. In other words, equality on the graph
structure is done strictly over links and the set of links attached to
each node.  Second, these graphs change over marker index; one or more
links may be in different configurations between distinct marker
points.  These graphs can be arbitrarily large, with an arbitrarily
large range of marker values over which links can change location, so
brute force equality testing at specific marker values quickly becomes
infeasible.  The computational problems are exacerbated when one
wishes to work over large collections of these, matching graph
structures and looking for patterns.
 
To set up this example, consider the graph shown in Figure
\ref{fig:example-graph}.  Snapshots of the graph $G$ are shown at
three marker values, $m_1 < m_2 < m_3$. Now consider $G(m_1)$ in
Figure \ref{fig:example-graph:a}. As the nodes and links all have
distinct labels, it can be represented by either listing the edges
connected to each node or the nodes connected to each edge as shown in
Tables \ref{tab:example-G-representations:a} and
\ref{tab:example-G-representations:b} respectively.  Note that the
structure of the graphs is uniquely described by the sets in the right
hand column of either table.  This allows us to test a graph structure
without considering labels on the nodes but only on the edges, as we
do for IBD graphs.  For equality testing purposes, this graph can be
represented exactly by first computing a hash over each connecting
set, then over the set of resulting hashes; this is essentially what
we do.

\newcommand{\evenspace}[1]{#1}
\newcommand{\tabheader}[1]{#1}

\begin{table}[tbh]
  \begin{centering}
    \centerline{
    \hss
      \subfloat[\label{tab:example-G-representations:a}]
      {
        \mbox{
          \begin{tabular}{|l|l|}
            \hline
            \tabheader{Node} & \tabheader{Edges} \\
            \hline
            $1$ & $A$ \\
            $2$ & $A,B,E$ \\
            $3$ & $B,C$ \\
            $4$ & $C,D$ \\
            $5$ & $D,E$ \\
            \hline
          \end{tabular}
        }
      }
      \hss
      \subfloat[\label{tab:example-G-representations:b}]
      {
        \mbox{
          \begin{tabular}{|l|l|}
            \hline
            \tabheader{Edge} & \tabheader{Nodes} \\
            \hline
            $A$ & $1,2$ \\
            $B$ & $2,3$ \\
            $C$ & $3,4$ \\
            $D$ & $4,5$ \\
            $E$ & $2,5$ \\
            \hline
          \end{tabular}
        }
      }
      \hss
      \subfloat[\label{tab:example-G-representations:c}]
      {
        \mbox{
          \begin{tabular}{|l|l|}
            \hline
            \tabheader{Node} & \tabheader{Edges} \\
            \hline
            $1$ & 
            $\evenspace{A}$, 
            $\evenspace{E}_\Ico{m_2, m_3}$ \\
            $2$ & 
            $\evenspace{A}$, 
            $\evenspace{B}$,
            $\evenspace{E}_\Ico{-\infty, m_2}$,
            $\evenspace{D}_\Ico{m_3, \infty}$
            \\
            $3$ &
            $\evenspace{B}$, 
            $\evenspace{C}$, 
            $\evenspace{D}_\Ico{m_2, m_3}$
            \\
            $4$ &
            $\evenspace{C}$,
            $\evenspace{D}_\Ico{-\infty, m_2}$,
            $\evenspace{E}_\Ico{m_3, \infty}$
            \\
            $5$ &
            $\evenspace{D}$
            $\evenspace{E}$ 
            \\
            \hline
          \end{tabular}
        }
      }
      \hss
    }
  \end{centering}
  \caption{\small{
      (a, b) Two representations of the graph $G(m_1)$ from
      Figure \ref{fig:example-graph:a}, by node (a) and by edge (b).
      (c) The same graph $G$ at all marker positions with 
      validity sets representing the differences in the graph 
      at marker values $m_1 < m_2 < m_3$.
    }}
  \label{tab:example-G-representations}
\end{table}

To extend this to dynamic graphs, we can associate validity
information with the components of the graphs.  Figure
\ref{fig:example-graph} shows $G$ at marker locations $m_1 < m_2 <
m_3$, with slight but significant changes between them.  Restricting
ourselves to looking only at the by-nodes representation in Table
\ref{tab:example-G-representations:a}, as the other is analogous and
this one is appropriate for IBD graphs, we can describe $G$ by Table
\ref{tab:example-G-representations:c}.  This produces a collection of
sets that varies by marker value; this example will be explored
further throughout this paper since working with dynamic collections
such as this one is the target application of our framework.

The paper is structured as follows.  In the next section we briefly
describe some related work, mostly involving innovative uses of
non-intersecting hashes.  In section \ref{sec:hashes}, we formalize
what we mean by a hash and describe a set of basic functions over them
with theoretic guarantees.  Then, in section \ref{sec:m-objects}, we
extend our theory of hashes to include marked hashes by describing
marker validity sets and the data structures we use to make marker
operations efficient.  Section \ref{sec:m-sets} details M-Sets,
our most significant contribution, along with available operations.
Finally, in section \ref{sec:examples}, we illustrate the flexibility
of our approach with several examples.  

\section{Related Work}
\label{sec:related}

Hash functions and related algorithms have seen numerous applications.
The unifying principle is that a short ``digest'' is calculated over
the message or data in such a way that changes in the data are
reflected, with sufficiently high probability, in the digest.  

Arguably the most widespread use of hash-like algorithms are with
check-sums and cyclic redundancy checks (CRCs) \citep{peterson11cyclic,
  maxino2009effectiveness, nakassis1988fletcher}. These are used to
verify data integrity in everything from file systems to Internet
transmission protocols, and are usually 32 or 64 bit and designed to
detect random errors.  The checksum is stored or transmitted along
with the data.  When the data is read or received, the checksum is
recalculated; if it doesn't match up with the original checksum, it is
assumed that an error occurred.

In cryptography, hashes, or ``message digests'', give a signature of a
message without revealing any information about the message itself
\citep{schneier2007applied}.  For example, it is common to store
passwords in terms of a hash; it is impossible to deduce what the
password is from the hash, but easy to check for a password match.
Much like a checksum, it is also used to ensure messages have not been
tampered with; as it is extremely difficult to produce different
messages that have the same digest.  It is this property that we
utilize in our approach.

Representing data by a hash is also common.  Hash tables, a data
structure for fast lookup of objects given a key, works by first
creating a hash of the key and using that hash to index a location in
an array in which to store the object \citep{cormen2001introduction}.
The hashes used in such tables are usually weak, as calculating the
hash is a significant efficiency bottleneck and the size of the lookup
array determines how many bits of the hash are actually needed --
usually not all.  Collisions -- distinct operations mapping to the
same hash -- may be common, so further equality testing is performed
to ensure the indexing keys match.  Thus such hash tables tend to be
relatively complicated structures.

Stronger hash functions usually produce hashes with 128 or more bits,
large enough that the probability of collisions is so low as to never
occur in practice.  Database applications often use such hashes to
index large files, as the non-existence of collisions greatly
simplifies processing \citep{silberschatz1997database}.  Similarly,
network applications often use such hashes to cache files -- files
having the same hash do not need to be retransmitted
\citep{karger1999web, wang1999survey, barish2000world}.  Often
cryptographic hash functions are used for this purpose; while slower,
they are computed without using network resources, so calculating them
is not the main efficiency bottleneck.  Furthermore, they are strong
enough that hash equality essentially guarantees object equality.

Our application extends several of these ideas, most notably the last
one.  We use strong hash functions to represent arbitrary objects in
our framework, assuming equalities among hashes are trustworthy.
However, our framework extends previous work in that it relies heavily
on several operations over hash values to reduce the information
present in collections down to a single hash that is invariant to
specified aspects of a process.  We present several theorems that
guarantee the summary hash is also strong. This allows us to reduce
computations that would be complex when performed over the original
data structures to simple operations over hashes while ensuring that
the results are accurate.

\section{Hashes}
\label{sec:hashes}

For our purpose, the hash function \Hash maps from an arbitrary object
or other hash to an integer in the set $\Hcal_N = \set{0,1,...,N-1}$
in a way that satisfies several properties.  First, such a function
must be {\it one-way}; i.e. no information about the original object
can be readily deduced from the hash (e.g. ``abcdef'' and ``Abcdef''
map to unrelated hashes).  In other words, having access to the output
of such a hash function is equivalent to having access only to an {\it
  oracle} function that returns true if a query object is equal to the
original and object and, with high probability, false otherwise
\citep{canetti1997towards,canetti1998perfectly}.

Under these requirements, \Hash can be seen as a discrete, uniformly
distributed random variable mapping from the event space -- arbitrary
objects or other hashes, etc. -- to $\Hcal_N$.  This form allows us to
assume $\Hash(\omega)$ has a uniform distribution on $\Hcal_N$ for an
arbitrary object $\omega$, a form useful for the proofs we give later
on.  Furthermore, the ``oracle'' property implies that the
distributions of two hashes are independent if the indexing objects
are distinct.

Second, \Hash must be strong, i.e.  collisions -- unequal objects
mapping to the same hash -- are extremely improbable.  Formally,

\begin{mathDefinition}[Strong Hash Function.] %
\label{def:strong-hash-function} %
A hash function \Hash mapping from an arbitrary object $\omega$ to a
an integer in $\Hcal_N = \set{0, 1, ..., N-1}$ is considered {\it
  strong} if, for $h_1 = \SC{Hash}(\omega_1)$ and $h_2 =
\SC{Hash}(\omega_2)$,
\begin{equation}
  \Pof{h_1 = h_2} = \fOO
  {1}{\omega_1 = \omega_2}
  {1 / N}{\omega_1 \neq \omega_2}
\label{eq:strong-hash}
\end{equation}
\end{mathDefinition}
\nid The idea is to set $N$ large enough (in our case around
$10^{38}$) that the probability of two unequal objects yielding the
same hash is so low as to never occur in practice.  However, in light
of the fact that collisions can theoretically occur with nonzero
probability, we denote inequality as $\nsim$ instead of $\neq$;
specifically, if $h_1 = \SC{Hash}(\omega_1) \nsim \SC{Hash}(\omega_2)
= h_2$, then $\Pof{h_1 = h_2} = 1/N$.

We may assume the existence of such a hash function, denoted here as
\Hash, which maps any possible input -- strings, numbers, other hashes
-- to a hash that satisfies definition \ref{def:strong-hash-function}.
This assumption is reasonable, as significant research in cryptography
has gone towards developing hash functions that not only satisfy
definition \ref{def:strong-hash-function}, but also prevent
adversaries with large amounts of computing power against deducing any
information about the original object \citep{goldreich2001foundations,
  schneier2007applied}.  These hash functions are widely available and
have open specifications; we use a tweaked version of the well known
\SC{md5} hash function as outlined in appendix \ref{sec:apx:hash}.

\subsection{Hash Operations}

Based on the existence of a hash function \Hash, and simple operations
on integers in $\Hcal_N$, we propose two basic operations to combine
and modify hashes.  The first is a way to summarize an unordered
collection of hashes by reducing it to a single hash that is sensitive
to changes in the hash value of any key in the original
collection. The second, to be used in nested function compositions, is
a way to scramble a reduced hash value so that it locks invariance
properties present earlier in the function composition.  In this
section, we formally describe these operations, which will later be
generalized to both marked hashes and then to collections of marked
hashes.

\subsubsection{Transformations}
\label{sec:transformations}

We now must formalize what we mean by a transformation in the testing
function context.  In our terminology, a transformation always applies
to the inputs of a testing function and is done without regard to the
hash values themselves.  For example, any reordering of the input
values is a valid transformation, but appending a precomputed string
to the label of an input object to cause its hash to be the special
null-hash -- all zeros -- is not (Note, however, that forming the
null-hash in this way is near-impossible in practice).  Formally,

\begin{mathDefinition}[Transformation Classes]
  \label{def:}
A transformation class for a testing function $\Tcal$ satisfies:
\begin{enumerate}[label=T\arabic*., ref=(T\arabic*)]
\item \label{lit:transformation:inputs} %
  Every $T \in \Tcal$ can be expressed as a transformation of the
  non-hash input objects.
\item \label{lit:transformation:nohashes} %
  No $T \in \Tcal$ takes account of the specific hash values produced
  by these objects.
\end{enumerate}
\end{mathDefinition} 
\nid Given these restrictions on transformation classes, we can now
formally define what we mean by invariance.
\begin{mathDefinition}[Invariance and Distinguishing]
  \label{def:invaraiance-and-distinguishing} %
  A function $f\,:\,\Hcal_N^n \mapsto \Hcal_n$ accepting a set of
  inputs $\mh = (h_1,h_2,...,h_n)$ is {\it invariant} under a class of
  transformations $\Tcal$ if $f(T\mh) = f(\mh)$ for all $T \in \Tcal$
  and for all $\mh \in \Hcal_N$.  Likewise, $f$ {\it distinguishes}
  $\Tcal$ if, for $T_1, T_2 \in \Tcal$, $f(T_1 \mh) \nsim f(T_2 \mh)
  \nsim f(\mh)$ unless $T_1\mh = \mh$, $T_2\mh = \mh$ or $T_1\mh =
  T_2\mh$.
\end{mathDefinition} %
\nid In other words, the output hashes change under distinguishing
transformations and are constant under invariant transformations.
With these formal definitions, we are now prepared to define atomic
operations that have specific and provable invariance properties.

\subsubsection{The Null Hash}
\label{sec:hash:nullhash}

We chose one value in our hash set, specifically 0, to represent a
{\it Null} hash.  This hash value, denoted as $\emptyset$, is used to
represent the absence of an input object.  It most commonly represents
the hash of an object that is outside its marker validity set; this
will be detailed more in section \ref{sec:m-objects}. As such, it has
special properties with the hash operations outlined in the next
section.

\subsection{Hash Operation Properties}
\label{sec:hash:properties}

We here propose two basic operations, \Reduce and \Rehash.  The first
reduces a collection of $n$ hashes, $h_1, h_2, ..., h_n$ for $n = 1,
2,...$, down to a single hash, while the second rehashes a single
input hash to prevent invariance properties from propagating further
through a function composition.  The key aspects of these operations
are what transformations over the inputs they are invariant under; we
describe these next.  We follow this with a brief discussion of the
implications of these results, before detailing the construction of
such functions in section \ref{sec:hash-reduction-operations}.

\begin{mathDefinition}[\Reduce]
  \label{def:hash:Reduce}

For the \Reduce function, with $h_0 = \Reduce(h_1,...,h_n)$, we have
the following properties:

\begin{enumerate}[label=RD\arabic*., ref={\textrm{RD\arabic*}}]
\item \label{lit:reduce:nullhash}
  {\it Invariance Under the null hash \NullHash}. %
  The output hash $h_0$ is invariant under input of the null hash
  \NullHash.  Specifically, 
  \begin{align}
    \Reduce(h_1, \NullHash) &= \Reduce(h_1) \\
    \Reduce(\NullHash) &= \NullHash
  \end{align}
\item \label{lit:reduce:single-mapping} %
  {\it Invariance Under Single Mapping}. %
  The output hash $h_0$ equals the input hash $h_1$ if $n = 1$.
  Specifically,
  \begin{equation}
    \Reduce(h_1) = h_1
  \end{equation}
\item \label{lit:reduce:inverse} %
  {\it Existence of a negating hash}. %
  There exists a negating hash, here labeled $-h_1$, that cancels the
  effect of an input hash in the sense that
  \begin{align}
    \Reduce(h_1, -h_1) &= \Reduce(\NullHash) = \NullHash
  \end{align}
\item \label{lit:reduce:order} %
  {\it Order Invariance}. %
  The output hash $h_0$ is invariant under different orderings of the
  input.  Specifically,
  \begin{equation}
    \Reduce(h_1, h_2) = \Reduce(h_2, h_1).
  \end{equation}
\item \label{lit:reduce:composition} %
  {\it Invariance Under Composition.}%
  The output hash $h_0$ is invariant under nested compositions of \Reduce.
  Specifically,
  \begin{align}
    \Reduce(h_1, h_2, h_3) &= \Reduce(h_1, \Reduce(h_2, h_3)) \\
    &= \Reduce(\Reduce(h_1, h_2), h_3)
  \end{align}
\item \label{lit:reduce:strength} %
  {\it Strength}. %
  The \Reduce function distinguishes all other transformations in the
  sense of definition \ref{def:invaraiance-and-distinguishing} (
  e.g. an input value is dropped or changed).
\end{enumerate}
\end{mathDefinition}

Two remarks are in order. First, property \ref{lit:reduce:composition}
allows us to expand all nestings of \Reduce to a single function of
hashes that are not the output of \Reduce operations.  For example, 
\begin{equation}
  \Reduce(h_1,\Reduce(h_2, \Reduce(h_3, h_4))) = \Reduce(h_1,h_2,h_3,h_4)
  \label{eq:574}
\end{equation}
The implication is that when we have a collection of input hashes
$h_1,h_2, ..., h_n$, which may or may not have come from a reduction
themselves, we can write their reduction out as a single reduction;
i.e.
\begin{equation}
  \Reduce(h_1,h_2,..., h_n) = \Reduce(h'_1, h'_2, ..., h'_{n'})
  \label{eq:reduce:expanding}
\end{equation}
for some hashes $h'_1, h'_2, ..., h'_{n'}$ that are not the result of
\Reduce.  

Second, property \ref{lit:reduce:inverse} allows us to remove elements
from a reduction once they are added, making the reduced hash
invariant under changes in whatever process produced the canceled
hash.  This property becomes especially useful later on when working
with intervals; the output hash varies as a function of a marker
value, and a hash $h$ valid on an interval $\Ico{a,b}$ of that marker
can be added once at $a$ and removed at $b$, with the net result being
that the output hash is only sensitive to $h$ on $\Ico{a,b}$.  

\begin{mathDefinition}[\Rehash]
\label{def:hash:rehash}

For the \Rehash function, we have only two properties, which we list
here.

\begin{enumerate}[label=RH\arabic*., ref=(RH\arabic*)]
\item \label{lit:rehash:nullhash}
  {\it Invariance Under the null hash \NullHash}.%
  The output hash $h_0$ is invariant under input of the null hash
  \NullHash.  Specifically, 
  \begin{align}
    \Rehash(\NullHash) = \NullHash 
  \end{align}
\item \label{lit:rehash:strength} %
  {\it Strength}. %
  The \Rehash function is strong in the sense of definition
  \ref{def:strong-hash-function}, in which the object space is
  restricted to hash keys.
\end{enumerate}
\end{mathDefinition}

\begin{figure}[t]
  \begin{centering}
    \includegraphics[width=\textwidth]{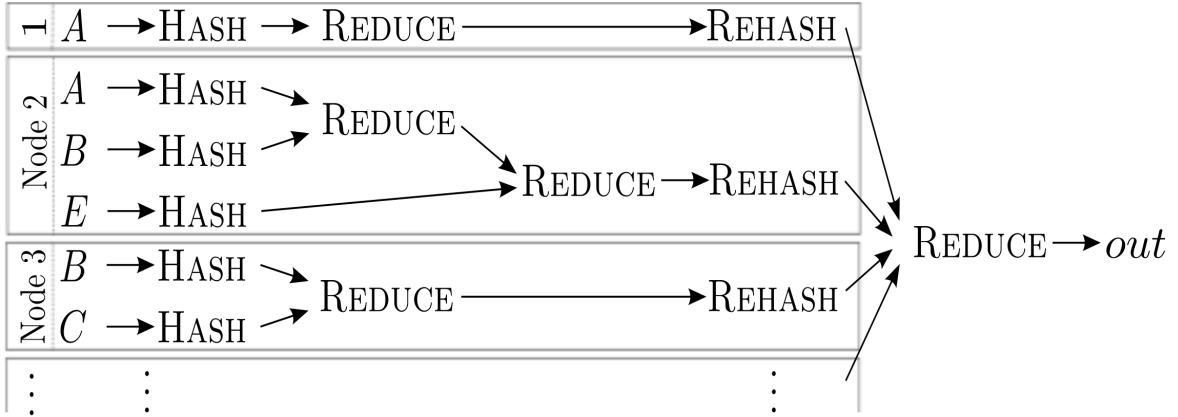}
  \end{centering}
  \caption{\small{Part of a composite function for testing equality,
      ignoring node labels, of the graph $G(m_1)$ in Figure
      \ref{fig:example-graph}. The part of the function for the
      first 3 nodes are shown.  For each node, the labels on the
      connected edges are reduced into one hash, and then the hashes
      representing the nodes are reduced to a final hash which
      represents the graph.  This final hash is invariant under
      changes in the node labels or orderings of the edges within each
      node, but is sensitive to any structural change in the graph.}}
  \label{fig:composition-invariance-example}
\end{figure}

The purpose of the \Rehash function is to freeze invariance patterns
from propagating through multiple compositions.  Returning to the
example IBD graph in Figure \ref{fig:example-graph}, consider the
testing function shown in Figure
\ref{fig:composition-invariance-example}. The function resulting from
chaining \Reduce and \Rehash together as shown is invariant under
changes in the node labels or orderings of the edges within each node,
but is sensitive to any structural change in the graph. This can be
proved by decomposing the nested \Reduce functions into a single
function of the first group; all the invariant relationships of this
single function are present in the original.  However, the final
reduce cannot be decomposed this way on account of the \Rehash
functions.

\subsection{Function Construction and Implementation}
\label{sec:hash-reduction-operations}

We now establish that functions $\SC{Reduce}$ and $\SC{Rehash}$
satisfying the appropriate properties exist.  Along with this comes a
requirement on $N$, namely that it is {\it prime}; this is required to
preserve the strength of the \Reduce function under multiple
reductions of the same hash key.

\subsubsection{Basic Operations}
\label{sec:lemmas}

Before presenting the \Reduce and \Rehash functions, we first present
two lemmas from elementary number theory.  These lemmas provide the
theoretical basis of the \Reduce function.
\begin{mathLemma}[]
  \label{lem:minus-commutes}
  Let $N$ be prime, and let $\oplus$ and $\otimes$ denote addition and
  multiplication modulo $N$, respectively. Suppose $a, b \in \Hcal_N$.
  Then the following equivalences hold modulo $N$:
  \begin{align}
    (a \otimes b) \oplus b &\equiv (a \oplus 1) \otimes b \label{eq:lem:distributivity} \\
    -\!(a \oplus b) &\equiv (-a \oplus -b) \\
    -\!(a \otimes b) &\equiv (-a \otimes b) \equiv (a \otimes -b)
  \end{align}
\end{mathLemma} %
\begin{proof} %
  The integers modulo $N$ forms an algebraic field with distributivity
  of multiplication over addition, so (\eqref{eq:lem:distributivity}) is
  trivially satisfied.  Furthermore,
  \begin{equation}
    -h \equiv (-1)\otimes h \equiv (N - 1) \otimes h \mod N
    \label{eq:620}
  \end{equation}
  Thus $-\!(a \oplus b) \mod N \equiv (-a \oplus -b) \mod N$.
  Similarly, multiplication is commutative, so 
  \begin{align}
    -\!(a \otimes b) &\equiv ((-1) \otimes a) \otimes b \equiv (-a \otimes b) \mod N \\
    &\equiv (a \otimes ((-1) \otimes b)) \equiv (a \otimes -b) \mod N 
    \label{eq:628}
  \end{align}
The lemma is proved.
\end{proof}

\begin{mathLemma}
  \label{lem:distributions}
  Let $N$ be prime, and let $X$ and $Y$ be independent random
  variables with distribution $\Un{\Hcal_N}$, i.e. uniform over
  $\Hcal_N$, and let $r$ be any number in $\Hcal_N$. Then
  \begin{align}
    C \oplus X &\sim \Un{\Hcal_N} \label{eq:lem:CpX} \\
    -X \mod N \equiv N - X &\sim \Un{\Hcal_N} \label{eq:lem:mX} \\
    X \oplus Y &\sim \Un{\Hcal_N} \label{eq:lem:XpY} \\
    r \otimes X &\sim \Un{\Hcal_N}, \label{eq:lem:CtX}
  \end{align}
  i.e. the above are all uniformly distributed on $\Hcal_N$.
\end{mathLemma} %
\begin{proof} %
For \eqref{eq:lem:CpX}, note that addition modulo a constant is a
one-to-one automorphic map on the hash space, thus every mapped number
is equally likely.  \eqref{eq:lem:mX} is similarly proved. To prove
\eqref{eq:lem:XpY}, note that $Y$ can be seen as a similar random
mapping; however, every possible mapping produces the same
distribution over $\Hcal_N$, so $X \oplus Y$ has the same distribution
as $X \oplus Y \C Y$, which, by \eqref{eq:lem:CpX} is uniform on
$\Hcal_N$.

For \eqref{eq:lem:CtX}, recall from number theory that $r$ has an
inverse $\mod N$ if and only if $r$ and $n$ are coprime, i.e. $\gca(r,
n) = 1$.  Thus if $N$ is prime, each $r$ in $\Hcal_N$ also indexes a
one-to-one automorphic map under $r \otimes X$, and the result
immediately follows.
\end{proof}

\subsubsection{Reduce}
\label{sec:638}

We are now ready to tackle \Reduce; if $N$ is prime, then addition
modulo $N$ satisfies all the required properties.  This operation is
similar to part of the Fletcher checksum algorithm, which uses
addition modulo a 16-bit prime for the reasons outlined in lemma
\ref{lem:distributions}.  

\begin{mathTheorem}[The \Reduce Function.] %
\label{th:reduce}
Suppose $N$ is prime.  Then the function $f\,:\,\calH_N^n \mapsto
\calH_N$ defined by
\begin{align}
  f(h_1, h_2, ..., h_n) 
  &= h_1 + h_2 + ... + h_n \mod N \\
  &= h_1 \oplus h_2 \oplus ... \oplus h_n,
\end{align}
where $\oplus$ denotes addition modulo $N$, satisfies properties
\ref{lit:reduce:nullhash}-\ref{lit:reduce:strength}.  \end{mathTheorem}
\begin{proof} Addition modulo $N$, with $N$ prime, forms an algebraic group, so
properties \ref{lit:reduce:nullhash}, \ref{lit:reduce:single-mapping},
\ref{lit:reduce:order} and \ref{lit:reduce:composition} are trivially
satisfied.  \ref{lit:reduce:inverse} is satisfied with $-h_1 \equiv N -
h_1 \mod N$.

To prove \ref{lit:reduce:strength}, it is sufficient to verify
equation \eqref{eq:strong-hash} in definition
\ref{lit:transformation:inputs}.  Let $h_0 = \Reduce(h_1,h_2, ...,
h_n)$, and let $k_0 = \Reduce(k_1, k_2, ..., k_m)$.  Without loss of
generality, by the previous properties, let the sequences be as
follows:
\begin{enumerate}[label=\arabic*., ref=(\arabic*)]
\item No hash in either sequence is the negative of another hash in
  that sequence.
\item $n \geq m$; if not, swap sequences.
\item There exists an index $q \in \set{1,2,..., m}$ such
  that $h_i = k_i$ for $i \leq q$ and $h_i \notin \set{k_{q+1}, ...,
    k_m}$ for $i = q+1, ..., m$.
\item $h_1 = k_1 = \NullHash$ (so $q \geq 1$, to make bookkeeping easier).
\end{enumerate}
Now suppose the two sequences are identical, so $n = m = q$.  Then
$h_0 = k_0$ and we are done; this satisfies the first part of equation
\eqref{eq:strong-hash}.  Otherwise, we can use lemma
\ref{lem:minus-commutes} to represent $h_0$ and $k_0$ as
\begin{align}
  h_0 &= Q \oplus (\alpha_1 \otimes H_1) 
  \oplus (\alpha_2 \otimes H_2) 
  \oplus \cdots 
  \oplus (\alpha_{n'} \otimes H_{n'}) \\
  k_0 &= Q \oplus (\beta_1 \otimes K_1) 
  \oplus (\beta_2 \otimes K_2) 
  \oplus \cdots 
  \oplus (\beta_{m'} \otimes K_{m'})
\end{align}
where $Q = \Reduce(h_1, h_2, ..., h_q)$ and $H_1, H_2, ..., H_{n'},
K_1, K_2, ..., K_{m'}$ are all independent and
$\alpha_1,...,\alpha_{n'}, \beta_1, ..., \beta_{m'}$ denote the
multiplicity of each hash.  Now it remains to show that $\Pof{h_0 \equiv
k_0 \mod N} = 1 / N$.  Now
\begin{equation}
  \Pof{h_0 \equiv k_0 \mod N} = \Pof{h_0 \oplus -\!k_0 \equiv 0}
  \label{eq:637}
\end{equation}
and, using lemma \ref{lem:minus-commutes} to distribute the minus
signs and eliminate the $Q$s,
\begin{align}
  h_0 \oplus -\!k_0 &= (\alpha_1 \otimes H_1) 
  \oplus (\alpha_2 \otimes H_2) 
  \oplus \cdots 
  \oplus (\alpha_{n'} \otimes H_{n'}) \nonumber\\
  &\quad\quad \oplus 
  (\beta_1 \otimes -\!K_1) 
  \oplus (\beta_2 \otimes -\!K_2) 
  \oplus \cdots 
  \oplus (\beta_{m'} \otimes -\!K_{m'}).
\end{align}
However, applying lemma \ref{lem:distributions} inductively gives that
the distribution of the above is uniform over $\Hcal_N$.  Thus
$\Pof{h_0=k_0}=1/N$.  \end{proof}

\subsubsection{Rehash}
\label{sec:732}
Now on to \Rehash, which is far simpler as it relies mainly on the
property of the hash function being strong.  The only extra work is to
ensure that the null hash is preserved.

\begin{mathTheorem}[\Rehash]
The function $\Rehash\,:\, \calH_N \mapsto \calH_N$ defined by 
\begin{equation}
  \Rehash(h) = \Reduce(\Hash(h), -\Hash(\emptyset)),
  \label{eq:384}
\end{equation}
satisfies properties
\ref{lit:rehash:nullhash}-\ref{lit:rehash:strength} while
distinguishing all other transformations.
\end{mathTheorem} %
\begin{proof} %
  Follows trivially from the properties of \Reduce and the assumption
  that \Hash is strong and one-way.
\end{proof} %

In this section, we have presented the fundamental building blocks
regarding hashes.  We now augment these hash values with validity
information that varies as a function of a particular parameter, here
called a marker value.

\section{Hashes and Keys}
\label{sec:m-objects}

We define a {\it key} as a hash value associated with a set of
intervals within which that hash, or the object it refers to, is
valid.  A key may represent an object in the data structure we wish to
design a testing function for, e.g. an edge in a graph that is present
only for certain marker values, or it may represent the result of a
process or sub-process.  At the marker values for which this key is not
valid, we assume its hash is equal to \NullHash.  To denote the hash
value of any key at a certain value $p$ of the parameter space, we use
brackets -- e.g. $h[p]$.

The set on which the hash value of a key is valid, which we call a
{\it marker validity set} or just {\it validity set}, is a sequence of
sorted, disjoint intervals of the form $\Ico{a_i,b_i} \subseteq
\Ico{-\infty, \infty}$.  A {\it marked} object is {\it valid} in each
of these intervals and {\it invalid} elsewhere.  Saying something is
{\it unmarked} is equivalent -- for bookkeeping reasons -- to saying
that it is always valid, i.e.  on the interval $\Ico{-\infty,
  \infty}$.

\section{Marked Sets}
\label{sec:m-sets}

The M-Set, a container of marked keys, is the most powerful
component of our framework.  It can be thought of as a collection of
marked objects, stored as representative keys, that permits easy
access to useful information about the collection.  The idea is that
one can express many algorithmically complicated processing tasks
involving dynamic data as simple operations on and between M-Set
objects.  Efficient operations on an M-Set include querying,
insertion, deletion, testing collection equality at specific marker
values or over the whole collection, union and intersection, and
extracting the collection of keys valid at specific marker values.

\subsection{Operations}
\label{sec:m-set:operations}

Available M-Set operations fall into five categories: {\it element
  operations} like insertion, querying, or modifying an element's
validity set; {\it hash and testing operations} like determining
whether two M-Sets are identical at marker $m$; {\it set operations}
such as union and intersection; {\it validity set operations} such as
extracting all hashes valid at a certain point; and {\it summarizing
  operations} which produce representative hashes from one or more
M-Sets.  Of these, operations in the first four categories are easily
explained; we present them in the next sections.  The summarizing
operation, which is key to the power of our framework, is presented in
section \ref{sec:m-set:summarize}.  A list of all these functions is
given in section \ref{sec:mset-operations}; the most powerful ones we
describe now.

\subsection{Summarizing Operations: \textnormal{\ReduceMSet }and \textnormal{\Summarize}}
\label{sec:m-set:summarize}

The natural generalization of \Reduce to M-Set objects, \ReduceMSet,
returns an M-Set containing the reduction of every key in the set.
Formally, for M-Sets $T_0$ and $T_1$, suppose $T_0 =
\ReduceMSet(T_1)$.  Then, for each marker value $m$, there is exactly
one key in $T_0$ valid at $m$, with that hash being the \Reduce of
every key in $T_1$ valid at $m$.  The M-Set is the appropriate output
of this function, as the resulting hash value varies arbitrarily by
marker value and thus cannot be expressed as a single key.  Because
such an M-Set has exactly one hash (possibly \NullHash) valid at each
marker value, we use the same bracket notation as keys to refer to
that hash value, e.g. $T[m]$.

Looking up the reduced hash of the M-Set at specific marker locations
-- \SC{HashAtMarker} -- is efficient to do without reducing the entire
set.  The main use for \ReduceMSet is thus to create a lookup of the
possible values of \Reduce in that set and when they are valid as
represented by the validity sets of the resulting keys. This can, for
example, be used to determine the set on which a dynamic collection is
equal to a given collection.

Just as \ReduceMSet summarizes the information in a collection of keys
by a single M-Set, so \Summarize reduces the information from one or
more distinct collections of M-Sets down to a single M-Set over which
computations can accurately and efficiently be done.  Changes in any
individual collection, as well as which collections are included, are
always reflected in the summarizing M-Set unless they fall under one
of the invariant properties (e.g. \NullHash does not affect the
outcome).

As such, \Summarize produces an M-Set $T$ in which one hash is valid
at each given marker position $m$.  For $T = \Summarize(T_1, T_2, ...,
T_n)$, the hash key valid at $m$, $T[m]$, is equal to
\begin{align}
  T[m] &= \Reduce(\Tcbrss{\Rehash(h)\;:\;\nonumber\\
  &\qquad h = \ReduceMSet(T_i)\text{ at } m,\text{ for } i = 1,2,...,n})
  \label{eq:1033}
\end{align}
Given our implementation of \ReduceMSet, described in the next section,
the \Summarize operation is very efficient and a central
building-block in our framework.

The summarizing operation is useful in that it allows us to
efficiently test equality of collections of M-Sets using the
operations designed for hashes.  For example, suppose we have two
summary M-Sets, $T = \Summarize(T_1, T_2, ..., T_n)$ and $U =
\Summarize(U_1, U_2, ..., U_n)$.  Given $T[p]$, the marker validity
set of the corresponding key in $U$, if any, gives the set in which
the collection of $T_i$'s at $p$ is equal to the collection of
$U_i$'s.  Likewise, \SC{MarkerUnion} over all objects in the
intersection of $T$ and $U$ gives the locations at which the two
collections are equal.

\subsection{Implementation}
\label{sec:m-set:implementation}

Internally, an M-Set is a combination of a hash table to store the
hashes and a skip-list-type structure that handles the bookkeeping
operations dealing with validity sets.  This latter structure
efficiently tracks the \Reduce at each marker value of all keys
present in the structure; this is key to making operations like
equality testing and summarizing efficient.

\subsubsection{Skip Lists for Markers}
\label{sec:1170}

To introduce the augmented skip-list for the \Reduce lookup, we first
describe a simpler version for holding marker information.  In a
skip-list, the values are stored in a single ordered linked list; this
allows for easy insertion and deletion, but by itself does not permit
efficient access.  To access them efficiently, there are additional
levels of increasingly sparse linked lists, each a subset of the
previous, with each node pointing forward and pointing down to the
corresponding node in the lower level.  When a new value is inserted
in the skip-list -- in our case a validity interval -- it also adds
corresponding nodes in the $L$ levels above it, where $L \sim
\Geometric{p}$.  The geometric distribution of the node ``heights''
means the expected size of level $L$ is $np^L$.  Overall, the expected
times for querying, insertion, or deletion is $\bigOof{\log n}$
\citep{papadakis1992average,kirschenhofer1994path,devroye1992limit}.

\begin{figure}[t]
  \begin{centering}
    \includegraphics[width=\textwidth]{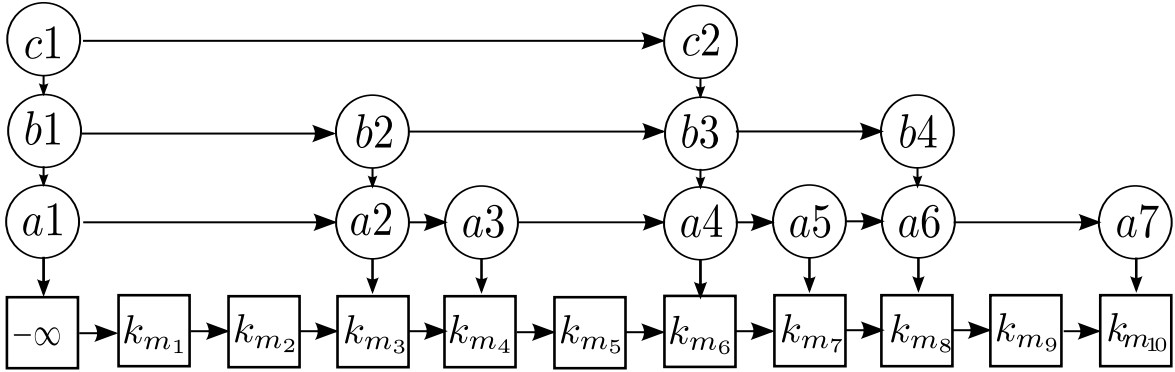}
  \end{centering}
  \caption{\small{A skip-list with 3 levels and 10 values. }}
  \label{fig:skip-lists}
\end{figure}
An example skip-list is shown in Figure \ref{fig:skip-lists}.  In this
skip-list, each marker location, denoted by $k_{m_1},...,k_{m_n}$, has
corresponding nodes in 0 to 3 levels above it.  The interval starting
values are stored in the nodes at each level.

Querying is done as follows.  Start at the first node in the highest
level, which is always at $-\infty$.  If a forward node exists and its
value is less than or equal to the query value, move forward;
otherwise, move down.  Repeat this until you're on the lower level and
cannot advance any farther; if this interval contains the query value,
that marker value is valid, otherwise it is not.  Finding locations
for insertion and deletion are analogous.  

\subsubsection{Internal \Reduce Lookup}
\label{sec:summary-hash-lookup}

We now present the data structure that comprises the second component
of an M-Set. This structure allows for calculating \Reduce, for a
given marker value, over all components of the entire hash collection
in logarithmic time while still allowing logarithmic time insertion
and deletion.  With this structure, we can also calculate \Reduce over
all validity set intervals values in time linear in the number of
distinct validity interval endpoints, producing a new M-Set of keys
with non-intersecting validity sets representing the different values
of \Reduce at various marker intervals.  These algorithmic bounds and
the associated algorithms will be formalized below.

The structure we propose is an augmented skip-list.  The leaf and node
values present in the tree correspond to the interval endpoints in the
validity intervals of any key present, i.e. the marker values where
the validity of any key changes.  This list is augmented to hold a
hash key in each leaf and each node.

The main idea for the leaf hashes is to track \Reduce over all valid
keys as a function of marker value $m$.  Stepping through the leaves,
starting with the hash value in the first leaf and updating it with
the leaf hash using \Reduce, yields \Reduce over all valid hashes at
each marker value.  This is done by including a hash at the beginning
of each of its validity interval and the negative of that hash at the
end of a validity interval.  Thus hash values are added and removed
from the overall reduced hash to maintain the value of \Reduce at each
marker value.  This allows us, additionally, to calculate the M-Set
produced by \ReduceMSet efficiently by simply stepping through the
leaves.

\begin{figure}[t]
  \begin{centering}
    \includegraphics[width=\textwidth]{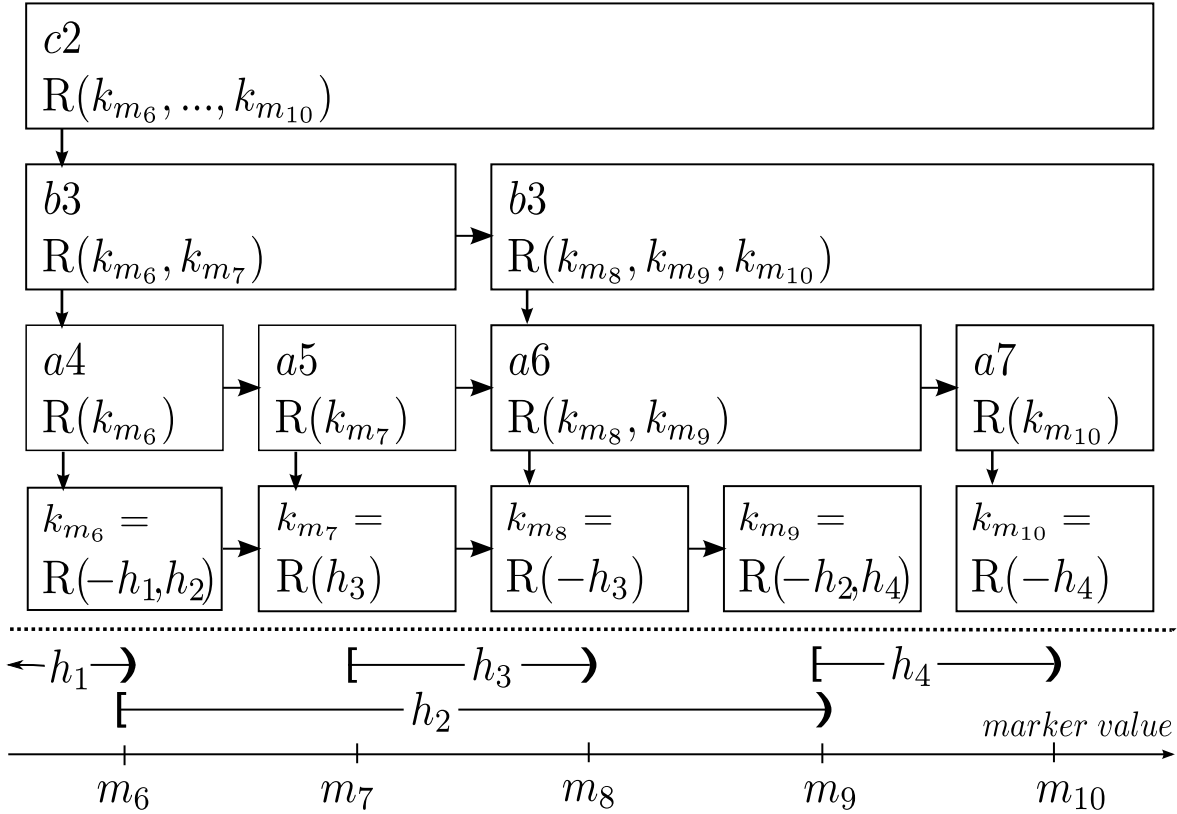}
  \end{centering}
  \caption{\small{The left part of the table in Figure
      \ref{fig:skip-lists} adapted to be an M-Set hash lookup
      structure holding hash information in the skip-list nodes. Hash
      keys, valid for certain marker intervals, are shown in the
      bottom, while the information stored in each node is shown below
      the node label. }}
  \label{fig:m-set-skip-list}
\end{figure}

Figure \ref{fig:m-set-skip-list} shows an example augmented
skip-list, the structure of which is taken from the right half of the
example skip-list in Figure \ref{fig:skip-lists}.  The leaves hold
hash values that are the reduction of hash values and/or negative hash
values.  At a given marker value, the value of \Reduce over all
previous leaves is the value of \Reduce over all currently valid keys.
For example, looking at the first three leaves,
\begin{align}
  &\Reduce(k_{m_6}, k_{m_7}, k_{m_8}) \\
  &\quad= \Reduce(\Reduce(h_1, h_2), \Reduce(-\!h_1, h_3), \Reduce(-\!h_3)) \\
  &\quad= \Reduce(h_1, h_2, -\!h_1, h_3, -\!h_3) \\
  &\quad= \Reduce(h_2)
  \label{eq:1139}
\end{align}
Formally, we maintain the following property:

\begin{mathProperty}[M-Set Marker Skip-List Leaf Hashes]
\label{prop:m-set:leaves}
For a given M-Set $T$ with skip-list $S$, let
\begin{align}
  L[m] &= \Set{h}{m = \ell\text{ for some $\Ico{\ell, u}$ in the validity set of $h$}} \\
  U[m] &= \Set{h}{m = u   \text{ for some $\Ico{\ell, u}$ in the validity set of $h$}}
\end{align}
Then for all leaves in $S$, define the hash value $r_0[m]$ at that
leaf to be
\begin{equation}
  r_0[m] = \Reduce(\Set{h}{h \in L[m]} \cup \Set{-h}{h \in U[m]})
  \label{eq:848}
\end{equation}
\end{mathProperty}

\nid Thus we can formally state the above.

\begin{mathTheorem}
Let $T$ be an M-Set with corresponding leaf nodes $r_0[\cdot]$. Let
$R[m] = \Reduce(\Set{r_0[m']}{m' \leq m})$.  Then
\begin{equation}
  R[m] = \Reduce(\Set{h[m]}{h \in T \text{ and } h \text{ is valid at }m}),
  \label{eq:862}
\end{equation}
Equivalently,
\begin{equation}
  R[m] = \Reduce(\Set{h[m]}{h \in T}),
\end{equation}
\end{mathTheorem}%
\begin{proof}%
On the marker intervals where a hash key is valid, its hash is
included in $R[m]$ exactly one time more than its inverse is included,
and it is included exactly the same number of times as its inverse at
all other values.  From property \ref{lit:reduce:inverse}, the summary
\Reduce hash at $m$ depends on a hash key if and only if that hash key
it is valid at $m$.  The equivalent formula for $R[m]$ follows
immediately from the fact that $h[m] = \NullHash$ if $h$ is not valid
at $m$; which does not change $R[m]$.
\end{proof}

As mentioned, the \Reduce of all leaf hash values whose associated
marker value is less than or equal to the given marker value $m$ is
equal to the \Reduce of all the keys in the M-Set valid at $m$.
However, storing the hashes in this way at the leaves is not enough to
efficiently compute \Reduce quickly over the full hash table at a
given marker value $m$, as it would require visiting every
change-point present that is less than $m$.  One might suggest storing
$R[m]$, the full value of \Reduce, at the leaves instead of just
$r_0[m]$, but then insertion and deletion would require time linear in
the number of marker points present in valid intervals, and this can
be arbitrarily large.

Our solution is to store a hash value summarizing blocks of $r_0[m]$
in the nodes at higher levels in the skip-list structure.  The idea is
that the hash value stored in the nodes is the reduction of all leaves
under it.  The presence of these hash values at the nodes allows us to
construct logarithmic time algorithms for querying, insertion and
deletion.  The idea is that we can include the reduction of large
blocks of nodes with a single operation as we travel down the
skip-list.

Formally, at the nodes, these hash values maintain the following
property:

\begin{mathProperty}[M-Set Skip-List Node Property] %
\label{prop:m-set:nodes}
Let $r_b[m]$ be a hash value at marker value $m$ in the $b$th level of
the skip-list with $b \geq 1$ ($b=0$ is the leaf level).  Let $m'$ be
the smallest marker value larger than $m$, possibly $\infty$, such
that there exists a node at level $b$ with marker value $m'$.  Then
\begin{equation}
  r_b[m] = \Reduce( \Set{r_{b-1}[m'']}{m \leq m'' < m'}).
  \label{eq:766}
\end{equation}
Equivalently,
\begin{equation}
  r_b[m] = \Reduce(\Set{r_0[m'']}{m \leq m' \leq m''})
  \label{eq:936}
\end{equation}
\end{mathProperty}
\nid In other words, the hash value of a node at level $b$, $b\geq 1$,
with marker value $m$ is the \Reduce over all nodes at level $b-1$
whose marker value is greater than $m$ and less than the marker value
of the next node at level $b$.  Property \ref{lit:reduce:composition}
of the reduce function -- invariance under composition -- means that
the hash value stored in a node is the \Reduce over all the leaf
values beneath it, i.e. reaching such a leaf requires passing through
that node.  This yields the equivalent formula \eqref{eq:936}.

In Figure \ref{fig:m-set-skip-list}, the hash at node $c2$ is the
\Reduce over the hash at nodes $b3$ and $b4$; the hash at node $b3$ is
the \Reduce over the hash at nodes $a4$ and $a5$, and so on.  The net
result of this is that the hash at each node is the \Reduce of all the
hash values beneath it.  

\begin{algorithm}[t]
  \SetKw{Step}{step}
  \dontprintsemicolon
  \caption{\SC{HashAtMarkerValue}}  
  \KwIn{M-Set $T$ and marker value $m$.}
  \KwOut{Hash Value $h$, the \Reduce over all hash objects in $T$ at $m$. }
  \Avspace
  \Assign{$h$}{$0$}
  \Assign{$n$}{First node of highest level of skip-list of $T$} 
  \Avspace
  \While{not at destination leaf}{
    \uIf{next node $n'$ has marker value $\leq m$}
    {
      \Assign{$h$}{$\Reduce(h, \text{hash at node $n$})$}
      \Assign{$n$}{$n'$}
    }
    \Else{
      \Assign{$n$}{node below $n$}
    }
  }
  \Avspace
  \Return{$h$}
  \label{algo:961}
\end{algorithm}

This allows us to calculate \Reduce at any marker value in logarithmic
time using Algorithm \ref{algo:961}.  This algorithm differs from the
regular skip-list query algorithm only in that it updates a running
hash as it traverses sideways. This means that at each point, the
current hash $h$ includes the reduction of all leaf nodes prior to the
current marker value, i.e. before moving forward, the reduction of all
leaf hashes between the current node and the next node is included in
\Reduce.  This last statement is sufficient to prove the validity of
the algorithm.

The algorithms for insertion and deletion are similar but involve more
detailed bookkeeping to handle the creation and deletion of nodes.
Apart from this, the only difference from Algorithm \ref{algo:961} is
that the hash at the node is updated when moving down, rather than
across; this preserves the invariant that the hash at a given node is
the $\Reduce$ of all the hash values stored under it.

\section{Example}
\label{sec:examples}

We now return to the motivating example, IBD graphs, given in section
\ref{sec:intro}.  The individuals in this case are edges, which are
assumed to be unique; the labels on the nodes are unidentifiable,
requiring any testing functions to be invariant to them.  

\begin{algorithm}[t!]
  \SetKw{Step}{step}
  \dontprintsemicolon
  \caption{IBD Graph Summarizing}
  \KwIn{An IBD Graph $G$.}
  \KwOut{$T$, an M-Set summarizing $G$; the hash of $T$ at a marker
    point $m$ is invariant under permutations of the node labels.}
  \Assign{$L$}{Empty list}
  \Avspace
  \For{Each node $n$ in $G$}{
    \Assign{$T$}{Empty M-Set}
    \For{Each edge $e$ attached to $n$ on interval $\Ico{t_1, t_2}$}{
      \Assign{$h$}{$\Hash\T{e}$}
      \Avspace
      \tcc{Set $\Ico{t_1,t_2}$ with key $h$ to be valid in $T$.}
      $\SC{AddValidRegion}(T,h,t_1,t_2)$
    }
    \Avspace
    \tcc{Append this new M-Set $T$ to our list.}
    \KwSty{append} $T$ \KwTo $L$\;
  } 
  \Avspace
  \tcc{The hash representation of the graph is the summary of all the
    graphs in $L$.} 
  \Assign{$T$}{$\Summarize(L)$}

  \Return$T$

  \label{algo:1420e}
\end{algorithm}

\begin{algorithm}[t!]
  \SetKw{Step}{step}
  \dontprintsemicolon

  \caption{IBD Graph Unique Elements}

  \KwIn{$S_1, S_2, ..., S_n$, M-Sets summarizing $n$ IBD graphs.}

  \KwOut{$L$, a list of $(h,\,\, mi, \,\, m)$ tuples giving a
    reference hash, an index, and a marker location denoting one
    instance of each unique graph in the original collection.}

  \Avspace

  \tcc{Form a single table of all unique graph hashes.}
  \Assign{$H$}{$\Union(\SC{KeySet}(S_1), \SC{KeySet}(S_2), ..., \SC{KeySet}(S_n))$}

  \tcc{Go through and find one index and marker value where each of
    these graphs occur.  $H$ tracks the graph hashes yet to be
    recorded.}

  \Assign{$L$}{Empty list}
  \For{$i = 1$ \KwTo $n$}{
    \Assign{$S$}{$\Intersection(S_i, H)$}
    \Avspace
    \ForEach{$h$ \KwSty{in} $S$}{

      \KwSty{append} $(h, i, \SC{VSetMin}(h))$ \KwTo $L$\;

      $\SC{Pop}(H, \SC{Key}(h))$\;
    }
  }

  \Return $L$

  \label{algo:1420f}
\end{algorithm}

The main idea is to represent each node as an M-Set with keys
representing edge labels.  The validity set on each key denotes when
that edge is attached to the node; this allows the structure of the
graph to change over marker location.  With each node represented this
way, the entire graph can represented as the summary of the node
M-Sets.  At each marker point, this computes a hash over each edge
within a node using \Reduce, rehashes the result to freeze invariants,
then computes a final hash over the resulting collections.  Per the
guarantees of \Reduce and \Rehash (definitions \ref{def:hash:Reduce}
and \ref{def:hash:rehash}), the resulting hashes of two graphs will
match if and only all the nodes have identical edges, which is true if
and only if the two graphs are equivalent (ignoring the completely
negligible probability of hash intersections).

Our first illustration, given in Algorithm \ref{algo:1420e}, simply
tests if two graphs are equal.  It also illustrates how to set up the
original graphs from a simple list-of-lists form.  Beyond this, we are
also be interested in all the unique graphs present in a collection of
node M-Sets.  Assuming these are summarized by $S_1, S_2, ..., S_n$ as
in Algorithm \ref{algo:1420e}, we can use algorithm \ref{algo:1420f}
to find a list of specific indices and marker locations that enumerate
the unique graphs.  Algorithms that need to be run, in theory, at each
marker value can instead be run only at this set of points.

\section{Experiments and Benchmarks}

To demonstrate the effectiveness of this approach, Table
\ref{tab:results} presents computation times on several real and
simulated IBD graph collections along with the savings incurred by
avoiding redundant operations.  The experiments were all run on an Intel
Xeon E5-4640 processor running at 2.40 GHz.  Recall that the
motivating computations to be run on the unique graphs (described in
section \ref{sec:intro}) can hours when run on a collection of these
graphs, so even a small reduction factor gives a significant time
savings and easily absorbs the preprocessing time shown here.  {\it
  Total Graph Configurations} is the number of potentially different
graphs over which a computation needs to be run.  On a single graph,
it is the total number of intervals on which there is no recorded
change in the graph; for multiple graphs, it is this factor summed
over all graphs.  {\it Unique Graphs} is the number of unique
configurations within this set; running computations only on each of
these is sufficient.

\newcommand\Rtab[1]{{\bf #1}}

\begin{table}[tbh]
\begin{small}
\begin{tabular}{|l|p{5em}|p{6em}|p{7.4em}|p{4em}|p{4em}|p{6em}|}
\hline 
\Rtab{Dataset} & \Rtab{Number of Graphs} & \Rtab{Individuals per Graph} &
\Rtab{Total Graph Configurations} & \Rtab{Unique Graphs} & \Rtab{Speedup Factor} & \Rtab{Computation Time} \\
\hline
Iceland-1 & 1000 & 95 & 155,612 & 150,290 & 1.04 & 2.18s \\
Iceland-2 & 1000 & 31 & 67,809 & 1,179 & 57.5 & 0.99s \\
Iceland-3 & 30000 & 31 & 1,616,028 & 1,376 & 1174.4 & 12.16s \\
\hline
fglhaps-7 & 1 & 7000 & 92,488 & 92,483 & 1.00005 & 10.39s \\
\hline
\end{tabular}
\end{small}
  \caption{\small{Result and processing times for Algorithm 
      \ref{algo:1420f} on several IBD graph datasets.}}
  \label{tab:results}
\end{table}

Table \ref{tab:results} shows results for four examples.  The three
Iceland datasets consist of IBD graphs realized conditionally on
marker data.  The marker data are simulated on a pedigree structure
described in \citep{glazner2012}.  Iceland-1 IBD graphs contain a full
set of 95 related individuals over 12 generations, while the graphs of
Iceland-2 and Iceland-3 are of a reduced set of individuals in the
last 3 generations for whom marker and trait data were assumed
available.  The fglhaps-7 example is a single IBD graph with $7000$
individuals and marker indexing from 1 to 140 million.  This graph
results from simulation of descent of a population of $7000$ individuals
over $200$ generations \citep{brown2012inferring}.

For the full Iceland graph on 95 individuals, Iceland-1, there is
little reduction in the number of graphs.  However, for the subset of
$31$ observed individuals for whom the probability $\Pof{Y_T | \mZ;
\gamma, \mGamma}$ must be computed (equation \eqref{eq:41AT}), there
is a greater than $50$-fold reduction even for only $1000$
realizations of the IBD graph.  When the number of realizations is
increased to $30,000$, the speedup is $3$ orders of magnitude, while
the time to process the IBD graphs increases only from 0.99s to 12.16s.
On the single graph of the fglhaps-7 example, 
there is little reduction from running the software,
since there are few marker intervals where the IBD graph is
repeated. However, this large is still processed by the software in a
relatively negligible 10.39s.

\begin{figure}[p]
  \begin{centering}
    \renewcommand*{\thesubfigure}{a}
      \subfloat{
        \includegraphics[width=0.81\linewidth]{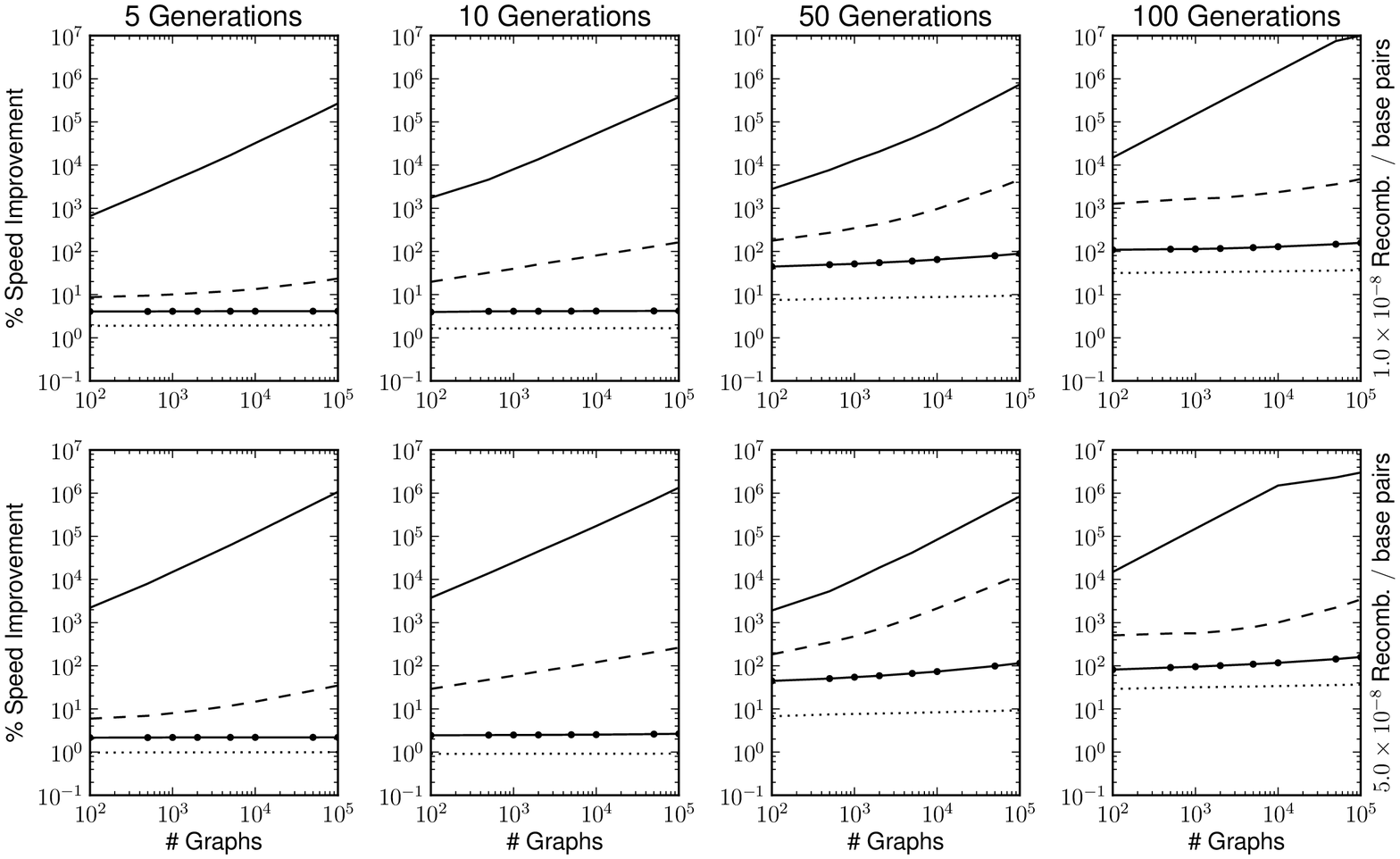}
      }
      \\
      \vspace{-15pt}
      \subfloat[\label{fig:1777:a}
      \small{Percentage speed improvement by redundant computations eliminated.}]
      {
        \includegraphics[width=0.7\linewidth]{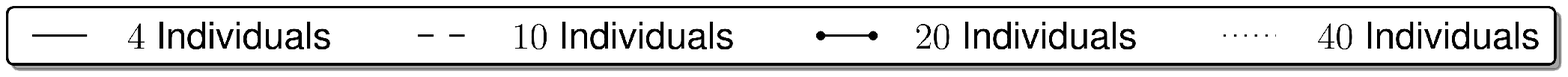}
      }
      \\
      \vspace{-12pt}
      \renewcommand*{\thesubfigure}{b}
      \subfloat{
        \includegraphics[width=0.97\linewidth]{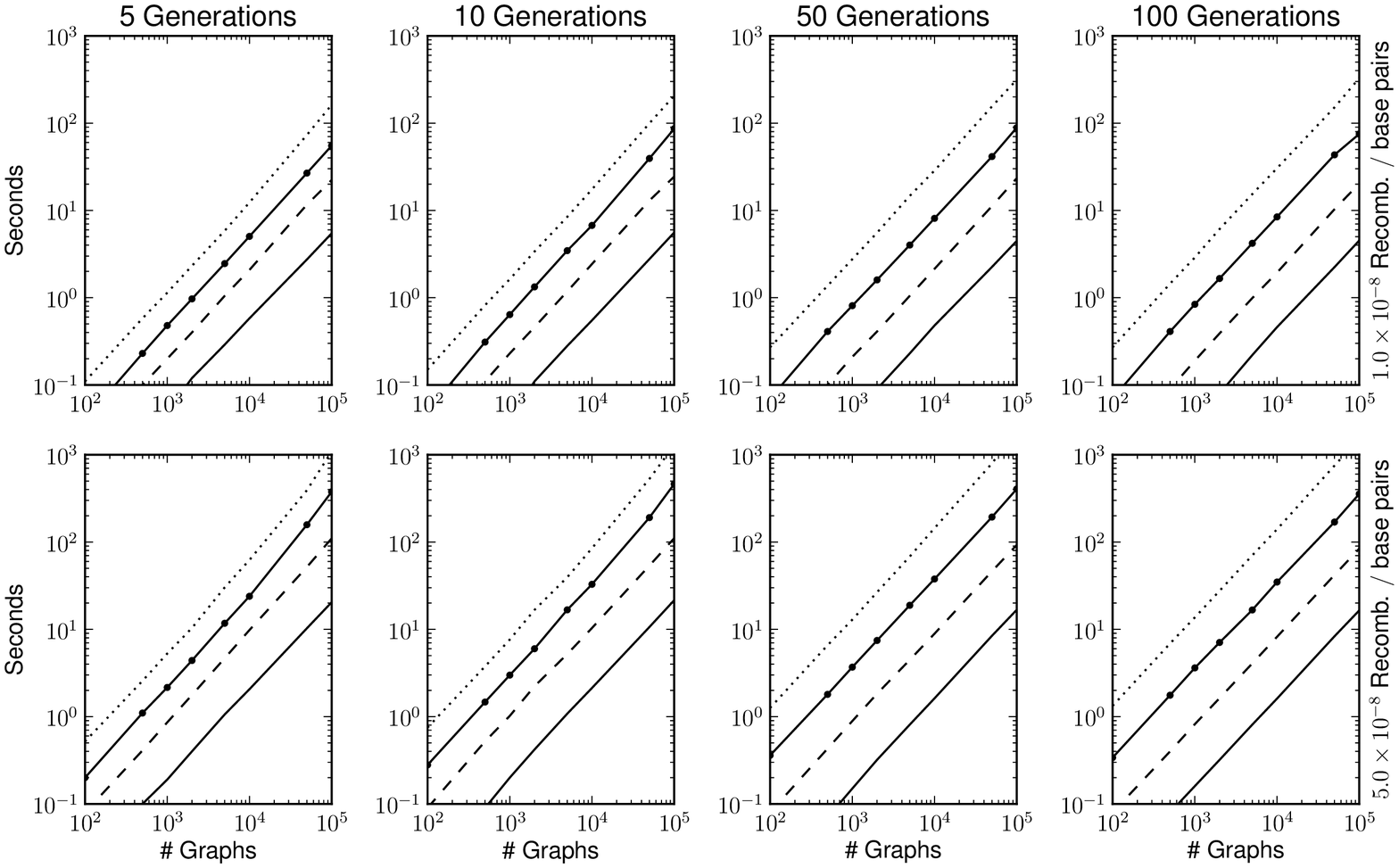} 
      }
      \\
      \vspace{-15pt}
      \subfloat[\label{fig:1781:b} 
      \small{Processing time required to compute equivalence classes.}]
      {
        \includegraphics[width=0.7\linewidth]{fig-legend-trimmed.eps}
      }
      \\
      \vspace{-4pt}
  \end{centering}
  \caption{\small{Results showing computation savings in
      simulated IBD graphs 
      generated from populations of 4,10,20,
      and 40 individuals (lines), with descent over 5, 10, 50, and 100
      generations (columns).  The recombination rates per generation
      are $1.0 \times 10^{-8}$ per base pair (top rows) and $5.0\times
      10^{-8}$ (bottom rows), with each individual a pair of chromosomes
      of length $10^8$ base
      pairs.  Results are shown for the IBD graphs 
      of the final generation ($x$-axes), for 
      sets of 100 to 100,000 realizations 
 }}
  \label{fig:results}
\end{figure}

In addition to this, Figure (\ref{fig:results}) shows the
computational results from simulation study of descent of chromosomes
of length $10^8$ base pairs over multiple population sizes, numbers of
realizations, number of generations, and recombination rates.  As can
be seen, for smaller population sizes, there is a substantial speed
improvement, often several orders of magnitude or more.  Furthermore,
as the number of realized IBD graphs in a collection increases,
disproportionally more redundancies are found, while the time required
to compute the equivalence classes scales linearly.  This indicates
that even if our method takes several minutes to run -- the most time
taken in these simulations -- it is always worthwhile.

These examples illustrate the power of our framework in working with
these types of dynamic data.  The advantage of M-Sets and the given
operations can be seen easily; many redundant operations can be
eliminated. Not surprisingly, these gains are the most substantial on
small graphs involving only a few individuals.  However, even in the
case where there is little reduction (e.g. fglhaps-7), the time taken
to process the equivalence classes is negligible relative to the rest
of the computations.  It should be noted that Iceland-2 and Iceland-3
showed the most dramatic reduction in processing time.  The Iceland
examples are those where the multiple IBD graphs are realizations
estimating a single true latent IBD graph, and are generated
conditional on genetic marker data.  The variation among graphs is
therefore much less than in the independent realizations of descent in
the other examples.  These Iceland examples demonstrate the
significant computational speed-ups that are possible in practice.

\section{Conclusion}
\label{sec:1633}

The representation of objects as hashes permits efficient set
operations, which in turn allows many testing algorithms to be
expressed in terms of these operations.  On more complex data,
summarizing and reduction operations allow data types with nested
representations to also work with this framework.  This is especially
true in the target structure, the IBD graph, in which otherwise
complex and slow tests can be expressed as simple and intuitive
operations.  Finally, we showed that real world operations can have
substantial speed improvements when using our framework to eliminate
redundant operations.  

The authors wish to thank Lucas Koepke for his contributions to the
code base, Steven Lewis for rigorously testing it, and Chris Glazner
for help with the experiments.  This open source library is freely available
online
at 
\url{http://www.stat.washington.edu/~hoytak/code/hashreduce}.

\appendix
\appsection{\Hash Function}
\label{sec:apx:hash}

The $\Hash$ function we use is CityHash \citep{CityHash}, which
produces a strong (though not cryptographic) 128 bit hash.  We map the
resulting hash to $\set{0,1,...,N}$, with the upper number chosen to
be prime.  In our case, we use $N = 2^{128} - 159$ as it is the
largest prime that can be represented by a 128 bit integer.

\appsection{Available Operations}

We here give a list of operations that are efficiently implemented in
our library.  

\subsection{Validity Set Operations}
\label{sec:mr:vset-operations}

To work with validity sets, we introduce several operations.  These
can be broken into two categories, operations that act directly on the
validity set of a key and operations that work between validity sets.
The former includes operations for constructing and manipulating a
validity set, testing whether a key is valid at a given marker value,
and iterating through a key's validity set intervals.  The latter
class implements set operations.  These operations all accept keys or
a marker validity sets as arguments and return a key or validity set
resulting from the respective operation.

\begin{description}[font=\normalfont,style=nextline,itemsep=0.1em]
\item[$\SC{IsValid}(X, m)$] Returns true if $m$ is a valid point in
  the validity set or hash object $X$ and false otherwise.
\item[$\SC{GetVSet}(h)$] Returns the validity set of a hash key $h$.
\item[$\SC{SetVSet}(h, M)$] Sets the validity set of a hash key $h$ to
  $M$.
\item[$\SC{AddVSetInterval}(X, a, b)$, $\SC{ClearVSetInterval}(X, a,
  b)$] Marks the interval $\Ico{a,b}$, $a < b$, as valid or invalid,
  respectively, in the validity set or hash object $X$.
\item[$\SC{VSetUnion}(X, Y)$, $\SC{VSetIntersection}(X, Y)$,
  $\SC{VSetDifference}(X, Y)$] Takes the set union, intersection, or
  difference between two validity sets or hash objects $X$ and $Y$,
  returning the result as a hash object if both $X$ and $Y$ are hash
  objects, and as a validity set otherwise.
\item[$\SC{VSetMin}(X)$] Returns the lowest valid marker value $m$.
\item[$\SC{VSetMax}(X)$] Returns the greatest marker value $m$ such
  that there are no valid regions greater than $m$.
\end{description}

\subsection{M-Set Operations}
\label{sec:mset-operations}

These operations are all efficiently implemented using the previously
described algorithms.

\subsubsection{Element Operations}
\label{sec:m-set:element}

\begin{description}[font=\normalfont,style=nextline,itemsep=0.1em]
\item[$\SC{Exists}(T, k)$] Returns true if a key with hash $k$ exists
  in $T$, and false otherwise.
\item[$\SC{ExistsAt}(T,k,m)$] Returns true if a key with hash $k$
  exists in $T$ and is valid at marker value $m$, and false otherwise.
\item[$\SC{Get}(T,k)$] Retrieves any key having hash $k$ from $T$.
\item[$\SC{Insert}(T,h)$] Inserts the key $h$ into $T$.
\item[$\SC{AddValidRegion}(T,h,t_1,t_2)$] Sets the region
  $\Ico{t_1,t_2}$ with key $h$ to be valid in $T$.  If $h$ is already
  present in the table, $\Ico{t_1,t_2}$ is set to be valid in that
  key's V-Set; otherwise, $h$ is given the V-Set $\Ico{t_1,t_2}$ and
  inserted into $T$.
\item[$\SC{Pop}(T,k)$] Removes any key having hash $k$ from $T$ and
  returns it.
\end{description}

\subsubsection{Hash and Testing Operations}
\label{sec:m-set:hash-and-testing-operations}

\begin{description}[font=\normalfont,style=nextline,itemsep=0.1em]
\item[$\SC{HashAtMarker}(T, m)$] %
  Returns the hash formed by \Reduce over all the keys valid at marker
  value $m$.
\item[$\SC{EqualAtMarker}(T_1, T_2, ..., T_n, m)$] %
  Returns true if all M-Sets $T_1, T_2, ..., T_n$ contain the same
  set of keys at marker $m$, and false otherwise.
\item[$\SC{EqualityVSet}(T_1, T_2, ..., T_n)$] %
  Returns a marker validity set indicating where all M-Sets are equal.
\item[$\SC{EqualToHash}(T, h)$] %
  Returns a validity set indicating the marker locations on which the
  reduction of $T$ is equal to the hash $h$.
\end{description}

\subsubsection{Set Operations}
\label{sec:m-set:set-operations}

\begin{description}[font=\normalfont,style=nextline,itemsep=0.1em]
\item[$\Union(T_1, T_2, ..., T_n)$] Returns an M-Set containing the
  union over all keys. For each hash value, the new marker validity
  set is the union of the validity sets of all keys having that key.
\item[$\Intersection(T_1, T_2, ..., T_n)$] Returns an M-Set containing
  the keys present in all input M-Sets, with the new validity set being
  the intersection of the originals' validity sets. Objects with no
  valid regions are discarded.
\item[$\Difference(T_1, T_2)$] Returns an M-Set containing all keys
  from $T_1$ with the validity sets from any corresponding hash in
  $T_2$ is removed.  Keys with empty validity sets are dropped.
\end{description}

\subsubsection{Marker Validity Set Operations}
\label{sec:m-set:validity-sets}

\begin{description}[font=\normalfont,style=nextline,itemsep=0.1em]
\item[$\MarkerUnion(T, M)$] Returns a new M-Set formed by all the
  keys in $T$, where the new validity sets are the union of the
  original and $M$.
\item[$\MarkerIntersection(T, M)$] Returns a new M-Set formed by all
  the keys in $T$, where the new validity sets are the intersection of
  the original and $M$.  Keys with empty validity sets are dropped.
\item[$\Snapshot(T, m)$] Takes a ``snapshot'' of the M-Set at a given
  marker value, returning an M-Set of all the hashes valid at that
  marker value.
\item[$\SC{KeySet}(T)$] Returns a new M-Set in which all keys in
  $T$ valid at any marker point in $T$ are returned as an unmarked
  set.  Equivalent to $\MarkerUnion(T, \Ico{-\infty, \infty})$.
\item[$\SC{UnionOfVSets}(T)$] Returns a validity set $M$ formed by
  taking the union of the validity set of every non-null key present
  in $T$.
\item[$\SC{IntersectionOfVSets}(T)$] Returns a validity set $M$ formed
  by taking the intersection of the validity set of every non-null key
  present in $T$.
\end{description}

\bibliographystyle{plainnat}
\bibliography{references}

\end{document}